# Geometry of Magnetic Fluctuations near the Sun from PSP

R. Bandyopadhyay[1], D. J. McComas[1]

[1]Department of Astrophysical Sciences, Princeton University, Princeton, NJ 08544, USA

**Abstract**
Solar wind magnetic fluctuations exhibit anisotropy due to the presence of a mean magnetic field in the form of the Parker spiral. Close to the Sun, direct measurements were not available until the recently launched Parker Solar Probe (PSP) mission. The nature of anisotropy and geometry of the magnetic fluctuations play a fundamental role in dissipation processes and in the transport of energetic particles in space. Using PSP data, we present measurements of geometry and anisotropy of the inner heliosphere magnetic fluctuations, from fluid to kinetic scales. The results are surprising and different from 1 au observations. We find that fluctuations evolve characteristically with size scale. However, unlike 1 au solar wind, at the outer scale, the fluctuations are dominated by wavevectors quasi-parallel to the local magnetic field. In the inertial range, average wave vectors become less field-aligned, but still remain more field aligned than near-Earth solar wind. In the dissipation range, the wavevectors become almost perpendicular to the local magnetic field in the dissipation range, to a much higher degree than those indicated by 1 au observations. We propose that this reduced degree of anisotropy in the outer scale and inertial range is due to the nature of large-scale forcing outside the solar corona.

**Introduction**

As the solar wind expands from the corona into interplanetary space, the solar magnetic field systematically gives up control of the coronal plasma, and in-situ dynamical processes, such as turbulence, begin to manifest. In particular, when the flow speed exceeds the local Alfvén speed (known as the Alfvén critical point), the magnetically controlled outer corona transforms into the kinetic-energy dominated solar wind, and large-scale, more isotropic turbulence starts to develop (DeForest et al. 2016a). This inner heliosphere solar wind, outside the Alfvén critical region, is sometimes called dynamically "young" or "nascent" solar wind. Parker Solar Probe (PSP), even during its first five orbits, provides the first opportunity to study the young solar wind with in-situ data (Fox et al. 2016; McComas et al. 2007).

Plasma turbulence, in the presence of a mean magnetic field, exhibits several types of anisotropies (Shebalin et al. 1983), including wavevector anisotropy and variance anisotropy (Matteini et al. 2020; Parashar et al. 2016). In the solar wind, the Parker spiral (Parker 1958) acts as a mean magnetic field; and therefore, both anisotropies are observed (Horbury et al. 2012; Oughton et al. 2015). Further, since the magnetic fluctuations are distributed over a broad range of length scales, the nature of anisotropy changes across scales. The spatial scales can be roughly grouped into three broad categories, from largest to smallest – outer scale or energy-containing scale, inertial range, and kinetic scale. Here, we study the nature of wave vector anisotropy near the sun across this wide range of scales.

We perform a slab+2D decomposition, which approximates the anisotropy in solar wind as the superposition of "slab" and "2-D" fluctuations (Bieber et al. 1996; Oughton 1993; Zank et al. 2017; Zank & Matthaeus 1992, 1993). The slab model assumes that wave vectors (**k**) are parallel to $\mathbf{B}_0$, and the 2-D model assumes that **k** is perpendicular to **B**$_0$. From incompressibility, the slab fluctuations are strictly transverse to **B**$_0$. Clearly, the slab+2D framework is highly idealized, with excitations restricted to modes either on the k$_\parallel$ axis or in the k$_\parallel$ = 0 plane. Thus, most of k-space is unpopulated which may appear as an oversimplification. Measurements and simulations show that turbulence power lies in between with a nontrivial distribution. Various models for this power distribution are discussed in the community (Galtier et al. 2000; Goldreich & Sridhar 1995; Shebalin et al. 1983; Sridhar & Goldreich 1994). Nevertheless, the slab+2D model has fared very well in matching observations, suggesting that it captures the important elements of the physics.

1 au solar wind observations indicate superposition of both slab and 2-D fluctuations from the "Maltese cross" plots (Matthaeus et al. 1990). Using the technique developed in Bieber et al. (Bieber et al. 1996) (which became known as the Bieber test), several studies [e.g., 19–24] have now established that 2-D fluctuations represent the dominant component (∼ 80% of the total energy) at this heliocentric distance. However, wind speed and latitude variation show some variability (Dasso et al. 2005; Smith 2003). The nature of slab and 2-D fluctuations close to the Sun are not well known (Adhikari et al. 2020; Cranmer 2018, 20). Here, we study the relative strength of the slab and 2-D components in the inner heliosphere from outer scale to the ion-kinetic range.

We analyze PSP data within the first five solar encounters (defined as the period when the spacecraft is within 0.25 au (Guo et al. 2021)). Following (Bandyopadhyay et al. 2020a; Chen et al. 2021; Parashar et al. 2020), we divide each encounter into intervals of about 8 hour duration, spanning several correlation lengths (see Table 1). We use publicly available Level-2 PSP/FIELDS flux-gate magnetometer (MAG; (Bale et al. 2016, 2019)) data for the correlation and inertial scale and Level-3 "moments" data from Solar Probe Cup (SPC; (Case et al. 2020; Kasper et al. 2016, 2019)) in the PSP/SWEAP archives, to obtain proton density and radial speed.

Table 1. Total data, number of intervals, time duration, and the total time used for energy-containing and inertial range analyses

| Data used | Number of intervals | Time duration of each interval | Total time analyzed |
|---|---|---|---|
| First 5 encounters | ~500 | 8 hours | ~4000 hours |

## Correlation-Scale Anisotropy

We begin by evaluating the two-dimensional correlation function R (r$_\perp$, r$_\parallel$) as a function of spatial scales parallel and perpendicular to **B**$_0$ to infer the nature of anisotropy at the outer scale. We rotate the magnetic field data into a mean field coordinate system for each interval (Belcher & Davis Jr. 1971, 197; Bieber et al. 1996). In this coordinate system, $\hat{\mathbf{z}}$ = **B**$_0$/|**B**$_0$| defines the mean-field unit vector, $\hat{\mathbf{y}} = -\hat{\mathbf{z}} \times \hat{\mathbf{R}}/|\hat{\mathbf{z}} \times \hat{\mathbf{R}}|$ where $\hat{\mathbf{R}}$ is the radial direction, and $\hat{\mathbf{x}} = \hat{\mathbf{y}} \times \hat{\mathbf{z}}$. The two-point correlation function is defined as $R(r) = \langle \mathbf{b}(\mathbf{x}) \cdot \mathbf{b}(\mathbf{x} + \mathbf{r}) \rangle$. Near the Sun, due to the low ratio of the Alfven speed to solar wind speed, Taylor's hypothesis may not always be valid, especially at electron scales (Huang & Sahraoui 2019). Through the first 5 encounters, Taylor's "frozen in" hypothesis is moderately maintained (Chasapis et al. 2020, 20); here we limit our analyses to MHD and ion kinetic scales. Therefore, we calculate the spatial lag $r$ from temporal lag $\tau$, using $r = V_{SW} \tau$ (Jokipii 1973; Taylor 1938). Here $V_{SW}$ is the mean solar wind flow speed in each interval. We employ a "Blackman-Tukey" technique (Blackman & Tukey 1958; Matthaeus & Goldstein 1982) to compute the correlation functions $R(r)$ for different values of the spatial lag $r$, with the maximum lag chosen as $1/5^{th}$ of the interval length. Then, we project the spatial separation $\mathbf{r}$ (radial, as it emerges from a frozen-in flow) onto a two-dimensional mesh spanned by $r_{\parallel}(= r \cos\theta$; where $\theta$ is the angle between $\hat{\mathbf{R}}$ and $\mathbf{B}_0$) and $r_{\perp}$ ($= r \sin\theta$) (Dasso et al. 2005, 200; Matthaeus et al. 1990; Osman et al. 2011; Weygand et al. 2009) and accumulate the estimates of $R(r)$ onto it.

Figure 1 shows the correlation function contour levels obtained from the first 5 encounters of PSP. The figure indicates both slab and 2-D fluctuations. The four-quadrant plot is produced by reflecting the data across
the axes from the first quadrant. In order to combine data from different intervals, we have normalized the fluctuations to the variance (e.g., Milano et al. 2001).

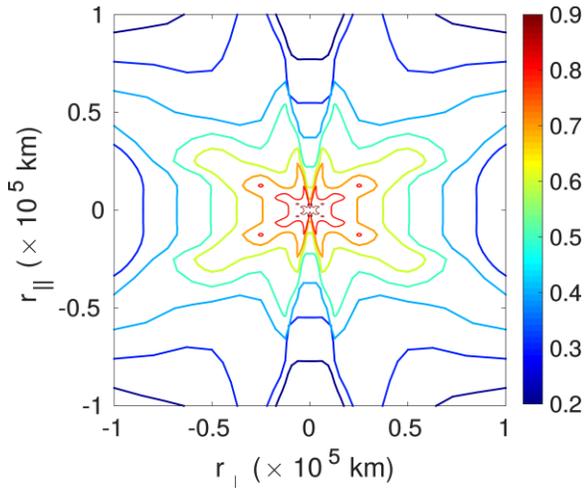

**Figure 1.** Two-dimensional correlation function from first 5 solar encounters of PSP.

Correlation lengths in the various angular bins (from ($r_{\perp}$, $r_{\parallel}$) bins using $\tan\theta = r_{\perp}/r_{\parallel}$) can be estimated as the values of $r$ where the decreasing function $R(r)$ drops by exponential ($R(0)/e$) (Smith et al. 2001). We consider all of the points in the angular bin $0° \leq \theta \leq 25°$ to calculate the parallel correlation length
$L_{\parallel}$ and those in $75° \leq \theta \leq 90°$ to calculate the perpendicular correlation length $L\perp$ fluctuations, respectively (Table 2).

**Table 2.** Parallel and perpendicular correlation lengths obtained from the two-dimensional correlation function (Fig. 1)

| $L_{\parallel}$ (km) | $L\perp$ (km) |
|---|---|
| $4.55 \times 10^4$ | $7.38 \times 10^4$ |

Our results show that the correlation length along the magnetic field is ~40% smaller than that in the perpendicular direction. This result indicates that the outer scale fluctuations are anisotropic, and the slab component is dominant at the outer scale. Although near-Earth observations have earlier shown that slab fluctuations can dominate or be comparable to 2D fluctuations at large scales, those observations were obtained only for fast wind (Dasso et al. 2005; Weygand et al. 2011). Slow wind exhibits a dominant 2D component at outer scale, and the ratio $L_{\parallel}/L_{\perp} > 1$ (Ruiz et al. 2010, 2011a, 2011b, 2014; Zhou & He 2021). PSP has sampled mostly slow wind during its first five solar encounters. Therefore, our result is in contrasts with 1 au observations. This suggests that the near-Sun slow wind is actually similar to fast wind at 1 au. This slow wind has had more time to evolve than the fast wind, so the two show significantly distinct properties at 1 au.

## Inertial-Range Anisotropy

Following Bieber et al. 1996 and Hamilton et al. 2008, we compute the ratio of energy associated with the slab ($E_s$) and 2-D ($E_2$) components from ratio of the spectra of the two perpendicular components, $P_{xx}$ and $P_{yy}$, and their power law index $q$.

$$\frac{P_{yy}}{P_{xx}} = \frac{k_s^{1-|q|} + r'\left(\frac{2}{1+|q|}\right)k_2^{1-|q|}}{k_s^{1-|q|} + r'\left(\frac{2|q|}{1+|q|}\right)k_2^{1-|q|}} \quad (1)$$

where $k_2 = 2\pi f/V_{SW}\cos\theta$ and $k_s = 2\pi f/V_{SW}\sin\theta$, $f$ is the frequency in the spacecraft frame, and $r' = E_s/E_2$. One obtains from Eq. (1)

$$r' = [tan\theta]^{1-|q|}\left(\frac{\frac{P_{yy}}{P_{xx}}-1}{|q|-\frac{P_{yy}}{P_{xx}}}\right)\left(\frac{1+|q|}{2}\right) \quad (2)$$

where $r = E_s/(E_s+E_2) = 1/(1+r')$ is the fraction of energy contained in the slab component within the selected frequency band.

We apply this formalism in two ways to estimate the energy partition. In the first analysis, we compute $r$ for each interval using Eq. (2) in the frequency interval 8-100 mHz to characterize the inertial range. We compute the mean value of the slab fraction $r$ for the database and provide the results in Table 3. Statistics are computed as described in (Hamilton et al. 2008).

**Table 3.** Averaged Parameters and Uncertainties in the Inertial Range.

| Power Law Indexes for Trace Spectra $\langle q \rangle \pm \sigma_q$ | Slab Energy Fraction $\langle r \rangle \pm \sigma_r$ |
|---|---|
| $-1.5646 \pm 0.0004$ | $0.546 \pm 0.009$ |

Compared to the well-known 80/20 ratio of 2D to slab energy, this reduced 2D power, observed in PSP, is rather surprising.

In the second analysis, we sum the computed ratio $P_{yy}/P_{xx}$ in bins of $\theta$ and then fit the results using Eq. (1) and the mean value of $q$ to obtain a least-squares fit value of $r$. This method has the disadvantage of using a single value for $q$ for all intervals but has the advantage of revealing the functional fit of Eq. (1).

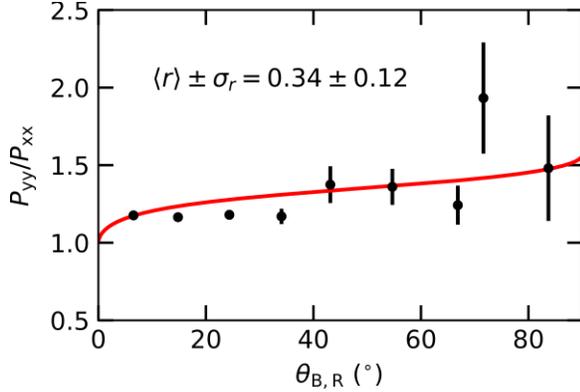

**Figure 2.** Distribution of $P_{yy}/P_{xx}$ binned by $\theta$ for inertial range. The error bars are computed as $\sigma/\sqrt{N}$, where $\sigma$ is the standard deviation and $N$ is the number of points in each bin.

Figure 2 plots the computed values of $P_{yy}/P_{xx}$ for 9 bins of $\theta$ for inertial range measurements for the first five PSP solar encounters, along with the minimum $\chi_r^2$ curve for the set of measured $P_{yy}/P_{xx}$. Using the average index $\langle q \rangle$ reported in Table 3. The resulting best fit value of $r$ is given in the figure. This analysis finds about 34% of energy in the slab components, which is still higher than the near-Earth observations of $r \sim 20\%$ in slab energy. Next, we repeat the same methodology in the kinetic range.

## Kinetic-Scale Anisotropy

To probe the nature of anisotropy in the dissipation range, we use SCaM data product of the FIELDS suite, which merges MAG and search-coil magnetometer (SCM) measurements (Bowen et al. 2020). This dataset resolves the ion kinetic range sufficiently above the noise floor (signal-to noise ratio > 5) (Huang et al. 2021). Due to a SCM anomaly in 2019 March, the full SCaM data are available only for the first encounter. Therefore, our kinetic-range analysis is limited to 2018 November 4 – November 7.

Table 3. Total data, number of intervals, time duration, and total time used for kinetic-range analyses

| Data used | Number of intervals | Time duration of each interval | Total time analyzed |
|---|---|---|---|
| 2018 November 4 -7 | ~2000 | 15 min | ~500 hours |

In the dissipation range, two distinct frequency bands are seen at the inner heliosphere from PSP (Huang et al. 2021): (i) an ion transition range ($f \sim$ 2–9 Hz), and (ii) an ion-kinetic range between the ion and electron scales ($f \sim$ 10–60 Hz). Here, we fit the ion kinetic range in the band $10 - 60$ Hz. We divide into 15-minute subintervals and calculate the energy partition to gather statistics. From Table 4, we find that the slab fraction is only ~10%.

**Table 4.** Averaged Parameters and Uncertainties in the Ion-Kinetic Range.

| Power Law Indexes for Trace Spectra $\langle q \rangle \pm \sigma_q$ | Slab Energy Fraction $\langle r \rangle \pm \sigma_r$ |
|---|---|
| $-2.109 \pm 0.003$ | $0.136 \pm 0.003$ |

We note here that the average index in the kinetic range is different than previous results at 1 au (e.g., Huang & Sahraoui 2019) and the first PSP perihelion (Duan et al. 2021; Huang et al. 2021). A possible reason is that here we selected only intervals where the index of $P_{xx}$ and $P_{yy}$ were similar and then calculated the average, while the other studies calculated the total power spectrum without any such restrictions.

Figure 3 shows the average values of $P_{yy}/P_{xx}$ in the ion-kinetic range for bins of $\theta$, along with the minimum $\chi_r^2$ curve. The resulting best fit value of $r$ is consistent with the value obtained from the average analysis and it is smaller than the slab fraction energy computed for the inertial scale in the previous section. This result is consistent with most previous studies at 1 au (Horbury et al. 2012; Podesta 2009; Sahraoui et al. 2010) and the inner heliosphere (Duan et al. 2021), showing that the fluctuations become more anisotropic at small scales.

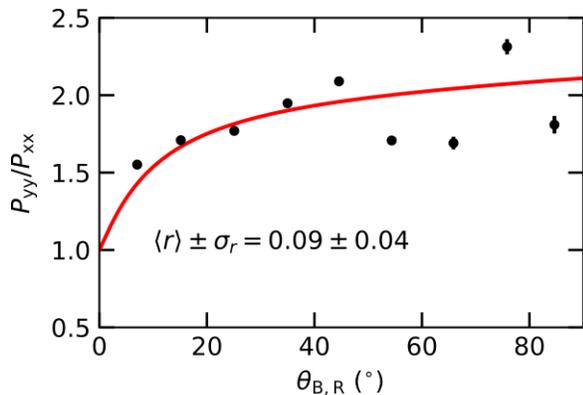

**Figure 3.** Distribution of $P_{yy}/P_{xx}$ binned by $\theta$ for ion-kinetic range. The error bars are $\sigma/\sqrt{N}$.

## Discussion

Simulations show that the Maltese-Cross type patterns with the simultaneous presence of slab and 2-D fluctuations do not arise naturally in a turbulent plasma, and such observations are possibly a consequence of either the initial condition or of averaging over different parcels of wind with each component (Ghosh et al. 1998). The PSP data, analyzed in the near-Sun environment, where the solar wind has had less time to develop, shows more about the origin of the slab and 2-D component (Duan et al. 2021; Zhu et al. 2020).

We propose the following scenario, which is summarized in Fig. 4, to explain our new results. In the sub-Alfvénic solar corona, below the Alfvén region, the plasma is essentially quasi-2D (Cranmer & Winebarger 2019, 2; Dmitruk et al. 2002; Matthaeus et al. 1999b, 1999a; Verdini et al. 2009, 200). Therefore, the increase of slab fraction in the outer scale and inertial range from PSP, compared to 1 au, suggests that there might be a transition region just beyond the Alfvén critical point, where the slab component is introduced and possibly amplified. The slab component is expected to increase due to the nature of mechanical forcing on the plasm parcels, presumably at large length scales. This "slab forcing" may be due to the shear, Kelvin-Helmholtz like roll ups, or waves (DeForest et al. 2016b; Ruffolo et al. 2020). It is possible that PSP is sampling outer part of this plasma.

Once the slab fluctuations are introduced at the forcing scales (outer scale), they eventually cascade to inertial and kinetic scale, but that takes time. This may explain why PSP observations show increased slab fraction at the outer and inertial scales, but reduced slab fraction at the kinetic range. Later, on the way to 1 au, most of the slab energy is converted to 2-D energy via anisotropic cascade at the inertial range. In the kinetic range, the 2-D fraction is preferably dissipated and, so the dissipation range is left with increased slab energy at 1 au. Later PSP orbits and future remote sensing and in-situ missions such as PUNCH and Solar Orbiter will shed more light on these explanations.

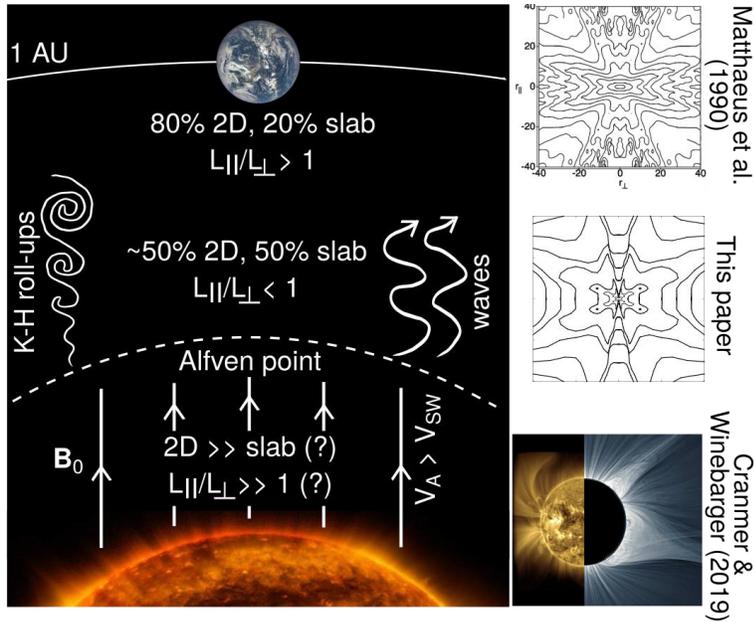

**Figure 4.** A schematic diagram (not drawn to scale) outlining the results of this paper and previous studies. The anisotropy information near 1 au were reported by (Bieber et al. 1996; Matthaeus et al. 1990) and subsequent papers. There are yet no in-situ data in the sub-Alfvenic corona so the nature of anisotropy there are not established, but remote sensing observations (Cranmer & Winebarger 2019) provide some clue. This paper analyzes near-coronal solar wind using PSP data.

The results of this paper provide new constraints on inner heliosphere plasma fluctuations that heat the solar wind and modulate the transport of SEP (Giacalone et al. 2006; Klein & Dalla 2017; Ruffolo et al. 2008). Quantification of the 2D and slab energy ratio is crucial for understanding Solar Energetic Particle (SEP) propagation close to the Sun (Bandyopadhyay et al. 2020b; Bieber et al. 1996; Droge 2003; McComas et al. 2016; Ruffolo et al. 2008).

The correlation length and its anisotropy are important parameters for understanding the dynamics of a plasma system (Bandyopadhyay et al. 2018, 2019) and for determining whether the parameters of numerical models match those of the space plasmas that they model (Adhikari et al. 2019; Usmanov et al. 2014, 2018). The direct estimation of parallel and perpendicular correlation scales also provides key inputs for solar wind models.

Turbulence-transport based solar wind models (Adhikari et al. 2015, 2017; Zank et al. 2017, 2018), often use 2D/slab decomposition. The results presented here are critical for these models. Finally, we expect that our results will eventually improve the day-to-day forecasting of space weather.

## Acknowledgements
We are deeply indebted to everyone that helped make the Parker Solar Probe (PSP) mission possible. This work was supported as a part of the PSP mission under contract NNN06AA01C. Parker Solar Probe was designed, built, and is now operated by the Johns Hopkins Applied Physics Laboratory as part of NASA's Living with a Star (LWS) program (contract NNN06AA01C). Support from the LWS management and technical team has played a critical role in the success of the Parker Solar Probe mission. All the data, used in this paper, are publicly available via the NASA Space Physics Data Facility (https://spdf.gsfc.nasa.gov/).